\documentclass[epj,nopacs]{svjour}
\usepackage{graphics}
\usepackage{amsfonts}          
\usepackage{amsmath}
\usepackage{amssymb}
\usepackage{graphicx}
\usepackage{flushend}
\usepackage{epstopdf}
\begin{document}
\title{Liquid droplets on a free-standing glassy membrane: deformation through the glass transition\\
}

\author{Adam Fortais\inst{1} \and Rafael D. Schulman \inst{1} \and Kari Dalnoki-Veress \inst{1,2}
\thanks{e-mail: dalnoki@mcmaster.ca}%
}                     
%
%
\institute{Department of Physics \& Astronomy,
McMaster University, Hamilton, ON, Canada \and Laboratoire de Physico-Chimie Th\'eorique, UMR CNRS Gulliver 7083, ESPCI Paris, PSL Research University, 75005 Paris, France.}
\date{June 22, 2017}
%
\abstract{In this study, micro-droplets are placed on thin, glassy, free-standing films where the Laplace pressure of the droplet deforms the free-standing film, creating a bulge. The film's tension is modulated by changing temperature continuously from well below the glass transition into the melt state of the film. The contact angle of the liquid droplet with the planar film as well as the angle of the bulge with the film are measured and found to be consistent with the contact angles predicted by a force balance at the contact line.
} 
\maketitle
\section{Introduction}
\label{introsec}

Elastocapillarity, the interplay between surface tension and elasticity, is of fundamental importance in many areas of research including microfluidics~\cite{Karpitschka2015}, self-assembly~\cite{Honschoten2010,Py2007,Paulsen2015,Patra2009,Huang2007}, substrate patterning~\cite{Style2013b,Shojaei-Zadeh2009,Chakrabarti2013}, wetting of fibers~\cite{Duprat2012,Protiere2013,Bico2004,Elettro2015b,Fargette2014}, and biological systems~\cite{Hazel2005,Grotberg2004,Campas2013}. One of the simplest geometries -- the contact angle a liquid drop makes with a soft solid -- has garnered a great deal of interest~\cite{Style2013b,Marchand2012,Hui2014,Style2012,Style2013a,bostwick2014,Pericet2008,Jerison2011,Nadermann2013,Hui2015,Schulman2015,Park2014}. In contrast to a droplet on a soft solid, the case of partial wetting of a liquid on a hard solid or a liquid on a liquid are well understood~\cite{deGennes2008}. A liquid droplet supported by an undeformable solid substrate will exhibit a contact angle, $\theta_y$, which can be calculated by performing a horizontal surface tension balance. This expression, which is known as Young's equation, is given by: $\gamma \cos{\theta_y}=\gamma_{sv}-\gamma_{sl}$, where $\gamma$ is the surface tension of the liquid, $\gamma_{sv}$ is the solid/vapor interfacial tension, and $\gamma_{sl}$ is the solid/liquid interfacial tension~\cite{deGennes2008,Young1805}. On a liquid substrate, the Laplace pressure of the droplet is able to deform the underlying liquid: for small droplets, where capillarity dominates gravity, a liquid lens is formed where the product of the curvature and the interfacial tension of both interfaces are equal (i.e. a Laplace pressure balance). Furthermore, at the contact line, the construction of a Neumann triangle in which the vertical and horizontal components of the surface tensions are simultaneously balanced allows one to determine the angles between the three interfaces~\cite{deGennes2008}. In the intermediate case in which substrates are soft, the droplet deforms the substrate at the contact line into a cusp with a length scale comparable to the elastocapillary length for bulk deformation $\gamma/E$ where $E$ is the elastic modulus~\cite{Style2013b,Hui2014,bostwick2014,Pericet2008,Jerison2011,Lester1961}. Microscopically, the contact angle between drop and substrate cusp satisfy a Neumann construction balancing the surface tension of the liquid and surface stresses between solid-liquid and solid-vapor interfaces. Macroscopically, the contact angle a drop larger than $\gamma/E$ makes with the planar, undeformed film satisfies Young's equation~\cite{Marchand2012,Style2012,Style2013a,Park2014}. However, for drops smaller than $\gamma/E$, these contact angles deviate from Young's equation~\cite{Style2013a,bostwick2014}.

In order to observe micrometer-scale elastocapillary deformations in the substrate, experiments are limited to soft materials with $E$ in the kPa range such that the elastocapillary lengths are on the order of a few~$\mu$m. An alternate approach is to utilize stiff materials in a compliant geometry, such as a thin free-standing film. In this geometry, bending of the thin membrane is of negligible cost in comparison to stretching, which permits macroscopic deformation of the film, while maintaining a sub-nanometer bulk elastocapillary length~\cite{Huang2007,Nadermann2013,Hui2015,Schulman2015,Davidovitch2013,Huang650,HUI2015116}.  

In this paper we consider the contact angles between liquid drops on thin, glassy free-standing films ($E\sim$ GPa) as a function of film tension. The tension is modulated by the thickness and temperature of the films, and the contact angles are measured continuously from well below the glass transition ($T_{g}$) of the film, into the melt state. We find that a Neumann construction accurately predicts the contact angle made between drops and the film.

\section{Experiment}  
\label{expsec}

Polystyrene (PS) with number averaged molecular weight,  $M_n = 451,000$ kg/mol, and polydispersity index of 1.10 (Polymer Source Inc.), was dissolved in toluene (Fisher Scientific, Optima grade) and spin cast onto freshly cleaved mica sheets (Ted Pella Inc.). The resulting films ranged in thickness, $h$, from 45~nm to 110~nm and were vacuum annealed for a minimum of 12 hours at 130~$^\circ$C to remove residual solvent and relax the polymer chains. After cooling to room temperature, films were cut with a scalpel blade into several smaller films and floated onto the surface of an ultra-pure water bath (18.2~M$\mathrm{\Omega}\cdot$cm). The floating films were subsequently transferred to 1~cm~$\times$~1~cm steel washers with a 3~mm diameter circular hole resulting in free-standing films. Additionally, one film from each sample was transferred to a silicon wafer for thickness measurements using ellipsometry (Accurion, EP3).

\begin{figure}
\includegraphics[width=1.0\columnwidth]{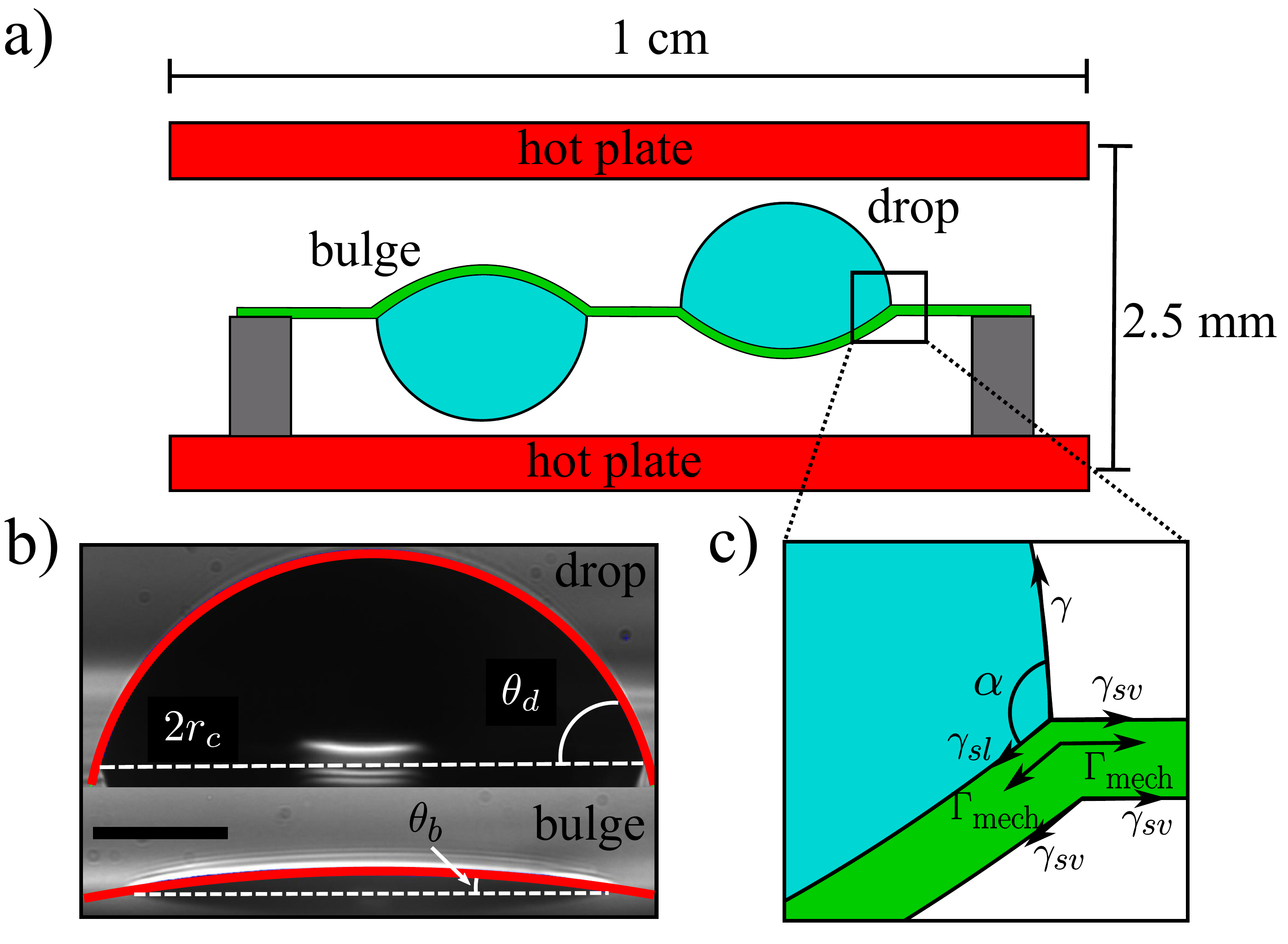}
\caption{a) Schematic of the side-view of glycerol drops on either side of a film between two temperature-controlled plates (not to scale). b) Microscope image of a drop (top) and bulge (bottom) with spherical cap fits (red curves). The contact diameters ($2r_{c}$) are indicated with dashed lines. Below  these lines is the reflection of the bulge/droplet by the film itself. The scale bar corresponds to 25~$\mu$m  c) Neumann construction accounting for interfacial and mechanical tension acting at the contact line. The interior angle $\alpha$ is the sum of $\theta_{b}$ and $\theta_{d}$. }
\label{fig1}
\end{figure}

The samples were placed in the experimental set-up shown schematically in figure~\ref{fig1}a and pre-annealed above the glass transition at 102~$^\circ$C for 2 minutes. This procedure relaxes any mechanical tension in the film, removes wrinkles which may have formed from transferring the film onto the washer, and ensures adhesion between the film and the steel washer.  After the temperature increases above the glass transition temperature, the film can flow to relax any pre-stress. Thus, at this point, wrinkles quickly relax and the film becomes a flat, stress-free liquid film. We ensure that due to the high viscosity at the annealing temperature no holes have formed in the liquid film which is inherently unstable to hole formation. In this melt state, the liquid film acquires a tension that is simply due to the surface tension of the two free interfaces. The sample was subsequently quenched into the glassy state to 65~$^\circ$C resulting in a taut film with uniform tension. Based on the thickness of the films and the speed of quench into the glassy state, we do not expect to observe any confinement effects \cite{PhysRevLett.95.025701}. Two glycerol (Caledon Laboratories Ltd.) drops, with volumes approximately 100 pL, were deposited on either side of the films as shown in figure~\ref{fig1}a. In this way, both the drop on the top side and the deformation in the film, due to the drop on the bottom side, could be imaged simultaneously from the side using an optical microscope. The Laplace pressure of the drop deforms the film, we refer to this deformed region of the film as the \emph{bulge}. The size of drops deposited were limited to contact radii between 25-100~$\mu$m, covering a combined contact area $<$~5\% of the film. Therefore, the deformation induced in the film by the droplets is perturbative to the overall size of the film. The lower limit in droplet size was imposed to reduce the effects of evaporation of the drop, while the upper limit assures that the additional stretching of the film is a perturbation to the tension. Lastly, all droplets were much smaller than the capillary length of glycerol so that gravity can be ignored.

The temperature of the hot plate was increased in 5$~^\circ$C increments up to 95~$^\circ$C at a rate of 90~$^\circ$C~/~min., remaining at each temperature for 5 minutes to allow the drops to reach an equilibrium contact angle before images were captured with a camera (QIMAGE QICAM Fast1394) attached to a horizontally mounted microscope objective. The sample is illuminated from behind. Since the sample is not in direct contact with the hot plates, a thermocouple (OMEGA HH506RA) was used to determine the actual temperature at the location of the sample. For measurements above 95~$^\circ$C, the temperature was increased directly to the desired temperature at a rate of 90~$^\circ$C/min, at which point images were captured at 15 second intervals, and contact angles were reported once they reached a constant value.

The contact radius ($r_{c}$) was measured directly from the microscope images, and the profiles were fit to spherical caps. Examples are shown in figure~\ref{fig1}b. From these fits, the radius of curvature of the drop and bulge ($R_{d}$ and $R_{b}$) were determined. The contact angle subtended by the spherical caps and the undeformed film away from the droplet ($\theta_{d}$, $\theta_{b}$) could be determined through the relation, $\sin(\theta_{d,b}) = r_{c}/R_{d,b}$, where the subscript $d$ or $b$ refers to either the droplet or bulge.

\section{Results and Discussion}
\label{ressec}

\subsection{Determining Tension}

It has been shown previously that the tension in the film can be determined by constructing a Neumann triangle at the contact line which includes both interfacial tensions and mechanical tension within the film~\cite{Nadermann2013,Schulman2015,HUI2015116,hui_planar_2015}. Figure~\ref{fig1}c is a schematic of the tensions acting at the contact line. 
By balancing these tensions in the vertical direction, an expression relating $\theta_{d}$ and $\theta_{b}$ to the tension in the film can be found:
\begin{equation}
\label{Eq1}
\frac{\sin\theta_{d}}{\sin\theta_{b}} = \frac{\Gamma_{\mathrm{mech}} + \gamma_{sv} + \gamma_{sl}}{\gamma},
\end{equation}
where $\Gamma_\mathrm{mech}$ is the mechanical tension within the film. Recently there has been significant interest in the difference between surface tension and surface stresses in soft solids \cite{andreotti_soft_2016,bostwick_elastocapillary_2014,andreotti_solid_2016,r._w._style_elastocapillarity:_2017}.Specifically, if the surface energy of the solid is strain-dependent, the surface stress is not equal to the surface energy. Since the strain-dependence of the surface energy of polymeric glasses has not yet been characterized, we make the simplifying assumption that the surface stresses equal the surface energies. In practice, it is not an important assumption because the mechanical tensions in the film are the dominant contribution to the total film tension. In addition, since our films are subject to very small strains ($<$1\%), the surface energies are not expected to deviate significantly from the unstrained values.

We note that rather than using the macroscopic contact angles, one can equivalently state that equilibrium requires that the Laplace pressure of the drop must balance that of the bulge:
\begin{equation}
\label{Eq2}
\frac{\gamma}{R_d}= \frac{\Gamma_{\mathrm{mech}} + \gamma_{sv} + \gamma_{sl}}{R_b}.
\end{equation}
By using the relation $\sin\theta_{d,b} = r_{c}/R_{d,b}$ and the Laplace pressure balance one recovers equation~(\ref{Eq1}). 

Equations~(\ref{Eq1}) and (\ref{Eq2}) show that by measuring the drop and bulge contact angles, or their radii of curvature, the total tension in the film (mechanical and interfacial) normalized by the surface tension of the liquid can be determined. This treatment relies on the assumption that the pretension of the membrane is unchanged by the addition of droplets when the deformation of the membrane induced by the droplet is a perturbation to the total area of the film. Additionally, because the drop is not pinned at the contact line, we assume the mechanical tension is uniform across the contact line. 

The assumption that the presence of small droplets results in a perturbative change to the tension of the membrane is in congruence with our observation that the measured tension does not change as additional droplets are placed on the film. Furthermore, because the bulge takes on the shape of a spherical cap we can assume the tension throughout the film below the drop is uniform. The accuracy of this technique has been confirmed elsewhere by comparison with micropipette deflection techniques \cite{Schulman2015}. 

By measuring the tension in films of various thicknesses at different temperatures, we can compare the measured values of the tension, to that predicted by using the material thermal expansion. In addition, we can further our understanding of the relationship between the droplet and bulge contact angles and the tension in the film.

\begin{figure}
\includegraphics[width=.95\columnwidth]{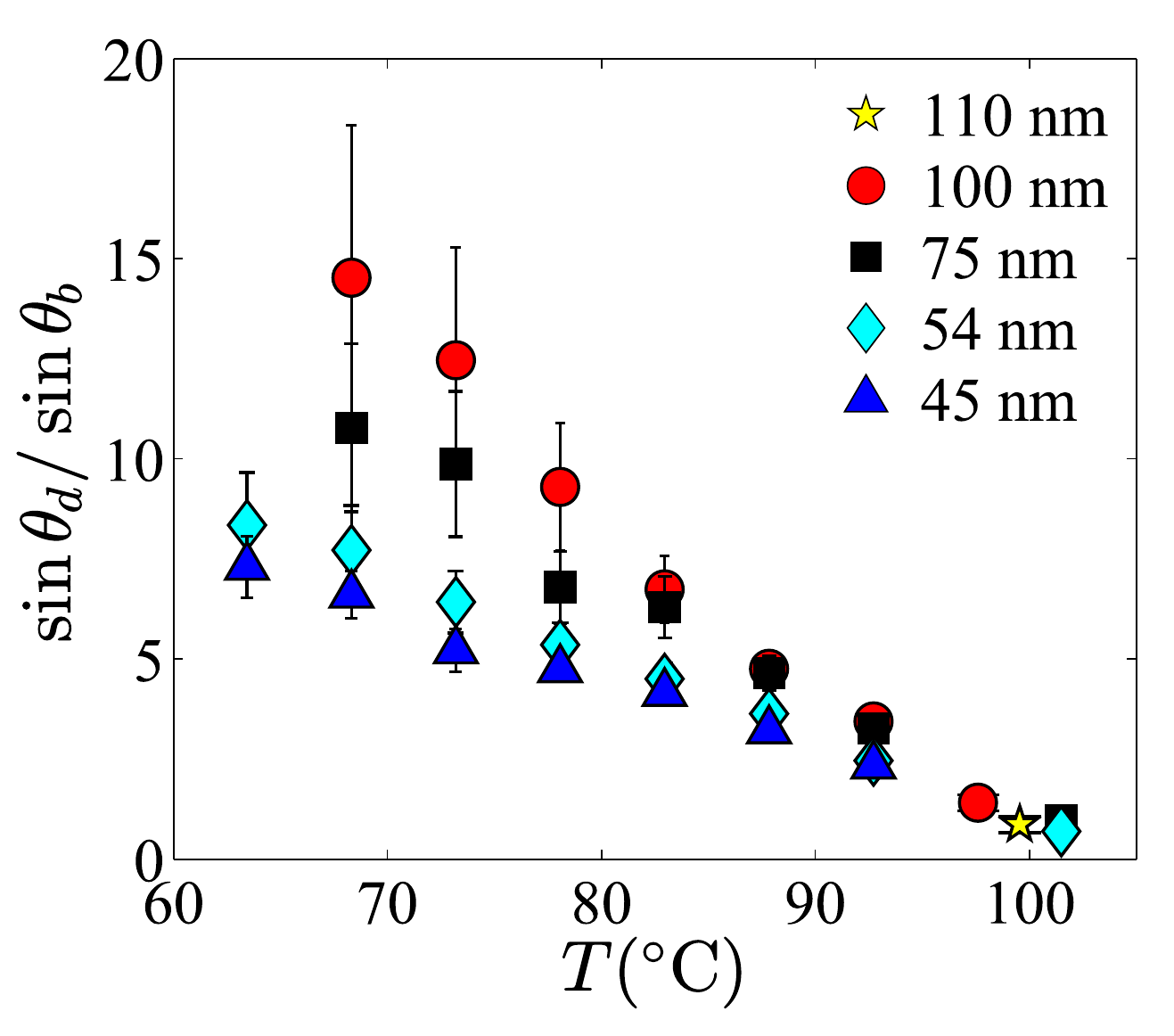}
\caption{The ratio $\sin\theta_{d}/\sin\theta_{b}$ as a function of temperature for various PS film thicknesses. This ratio is equivalent to  the tension of the membrane in contact with the droplet normalized by the surface tension of glycerol as given in eq.~(\ref{Eq1}).}
\label{sind_sinb}
\end{figure}

Guided by equation~(\ref{Eq1}) we plot the experimental observable quantity $\sin\theta_{d}/\sin\theta_{b}$ as a function of temperature for various film thicknesses in figure~\ref{sind_sinb}.  We recognize this ratio as being equivalent to the normalized tension of the membrane in contact with the droplet which includes the mechanical tension of the membrane and the relevant interfacial tensions [eq.~(\ref{Eq1})].  As seen in fig.~\ref{sind_sinb}, the normalized tension decreases with increasing temperature and increases as the film thickness increases. These trends are explained next.

The decrease in the normalized tension is due in part to a small change in the interfacial tensions with temperature, but dominated by a reduction in the mechanical tension of the membrane with increasing temperature. The changes in the mechanical tension as a function of temperature are caused by the difference in the thermal expansion of the PS films and the steel washer to which the film is affixed. The change in mechanical tension can be understood as follows. First, the film is taken above the glass transition temperature into the melt state to equilibrate. The tension in the film is then only due to the surface tension of  the two free interfaces. As the film is cooled below the glass transition, both the glassy PS  and the steel washer contract, however since the expansion coefficient of PS is greater than that of stainless steel, the cooling results in a uniform strain $\epsilon$ which increases with decreasing temperature. The strain is given by $\epsilon = (c_{PS}-c_{SS}) \Delta T$ where $c_{PS} = 70 \cdot 10^{-6}$~K$^{-1}$ is the linear expansion coefficient of PS and $c_{SS}$ is the linear expansion coefficient of stainless steel and depends on the alloy composition but is typically within the range $(14 \pm 3) \cdot 10^{-6}$~K$^{-1}$~\cite{Brandrup1999,gale2003smithells}. 
Lastly, the dependence on the film thickness in figure~\ref{sind_sinb} is easily understood from the fact that the mechanical tension in the membrane is proportional to the film thickness: while the strain due to differential thermal expansion is independent of the film thickness, the tension depends on the product of the strain and film thickness.

\begin{figure}
\includegraphics[width=1.0\columnwidth]{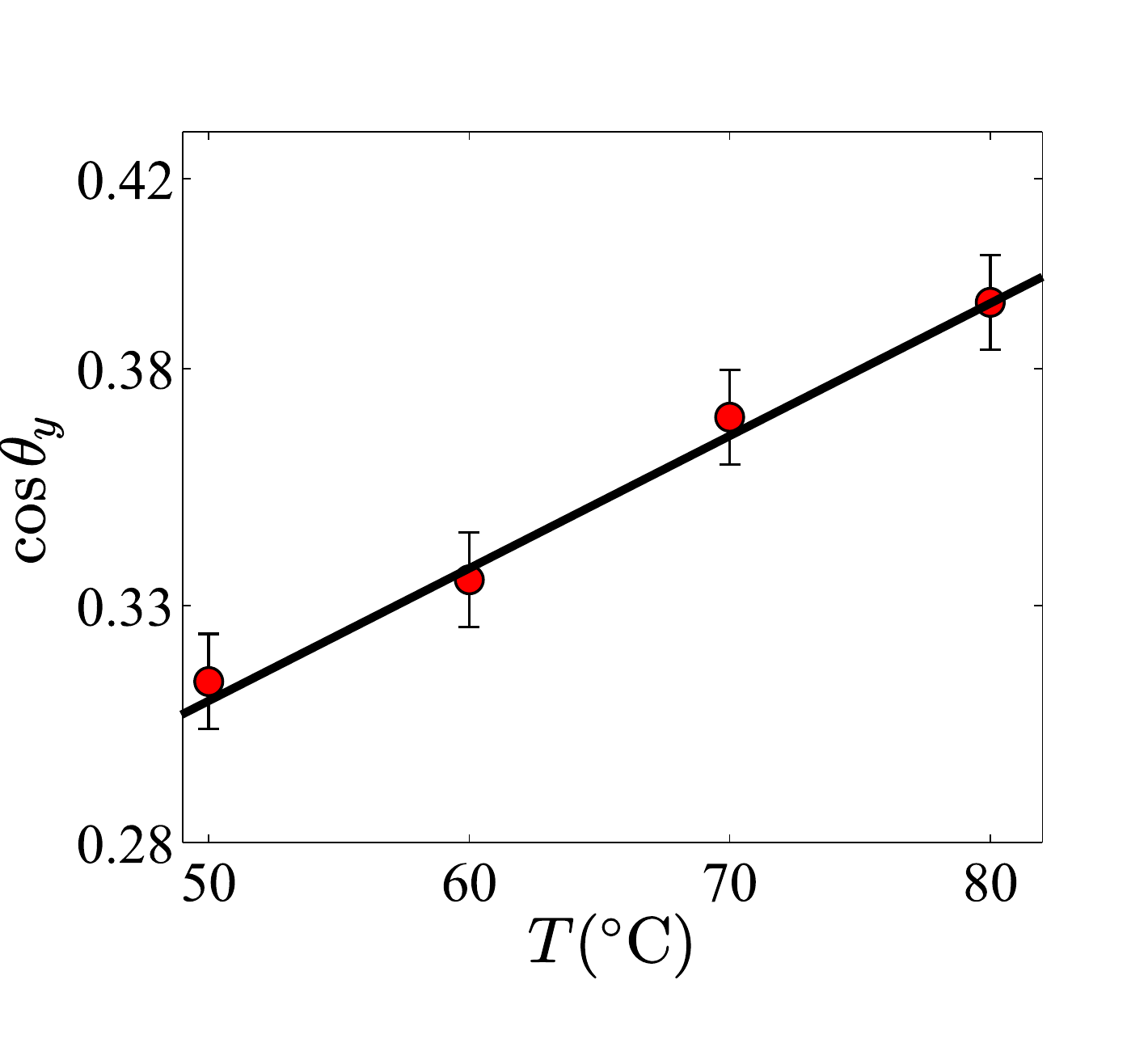}
\caption{Cosine of  experimentally determined Young's angle of glycerol droplets on a PS substrate as a function of temperature. The solid black line, $\cos \theta_y = aT+b$,  is a best fit to the data. Parameters $a$ and $b$ are given in the text. }
\label{youngs}
\end{figure}

In order to fully probe the mechanical tension,  $\Gamma_{\mathrm{mech}}$,  as a function of temperature, we can also take into account the variations in the interfacial tensions with temperature, $\gamma$, $\gamma_{sv}$, and $\gamma_{sl}$;  in addition to the dominant dependence of the film tension on thickness and the differential thermal expansion discussed above. While $\gamma(T)$ and $\gamma_{sv}(T)$ can be found in the literature, the interfacial tension for PS and glycerol can be found though Young's equation, $\gamma_{sl}=\gamma_{sv}-\gamma \cos \theta_y$. Using eq.~(\ref{Eq1}) and the Young's equation, we can write the mechanical tension in terms of accessible quantities as:
\begin{equation}
\label{eq:mech1}
\Gamma_{\mathrm{mech}} = \frac{\sin\theta_{d}}{\sin\theta_{b}}\gamma - 2\gamma_{sv} + \gamma \cos \theta_{y},
\end{equation}
where we stress that all quantities are dependent on temperature. The surface tension of glycerol is $\gamma=63-0.0598(T-20^{\circ}$C)~mN/m~\cite{Lide2006} and for PS, $\gamma_{sv} = 40.7 - 0.072(T-20^{\circ}\mathrm{C})$~mN/m~\cite{Wu1970}. We have measured the temperature dependence of Young's angle $\theta_{y}$ for glycerol atop a non-compliant film of PS supported on a rigid substrate (Si) as shown in  fig.~\ref{youngs}. The data for $\theta_{y}(T)$ are well described by the linear approximation $\cos \theta_{y} =aT + b$, where ${a = (0.27 \pm 0.04)\cdot 10^{-2} \ ^{\circ}\mathrm{C}^{-1}}$ and ${b = (17 \pm 5)\cdot 10^{-2}}$. Thus, we can write the mechanical tension as, 
\begin{equation}
\label{eq:mech2}
\Gamma_{\mathrm{mech}}(T) = \gamma(T) \left(  \left. \frac{\sin\theta_{d}}{\sin\theta_{b}}\right|_T+ aT + b \right) - 2\gamma_{sv}(T) ,
\end{equation}
where all temperature dependencies are explicitly indicated.

\begin{figure}[t!]
\includegraphics[width=.95\columnwidth]{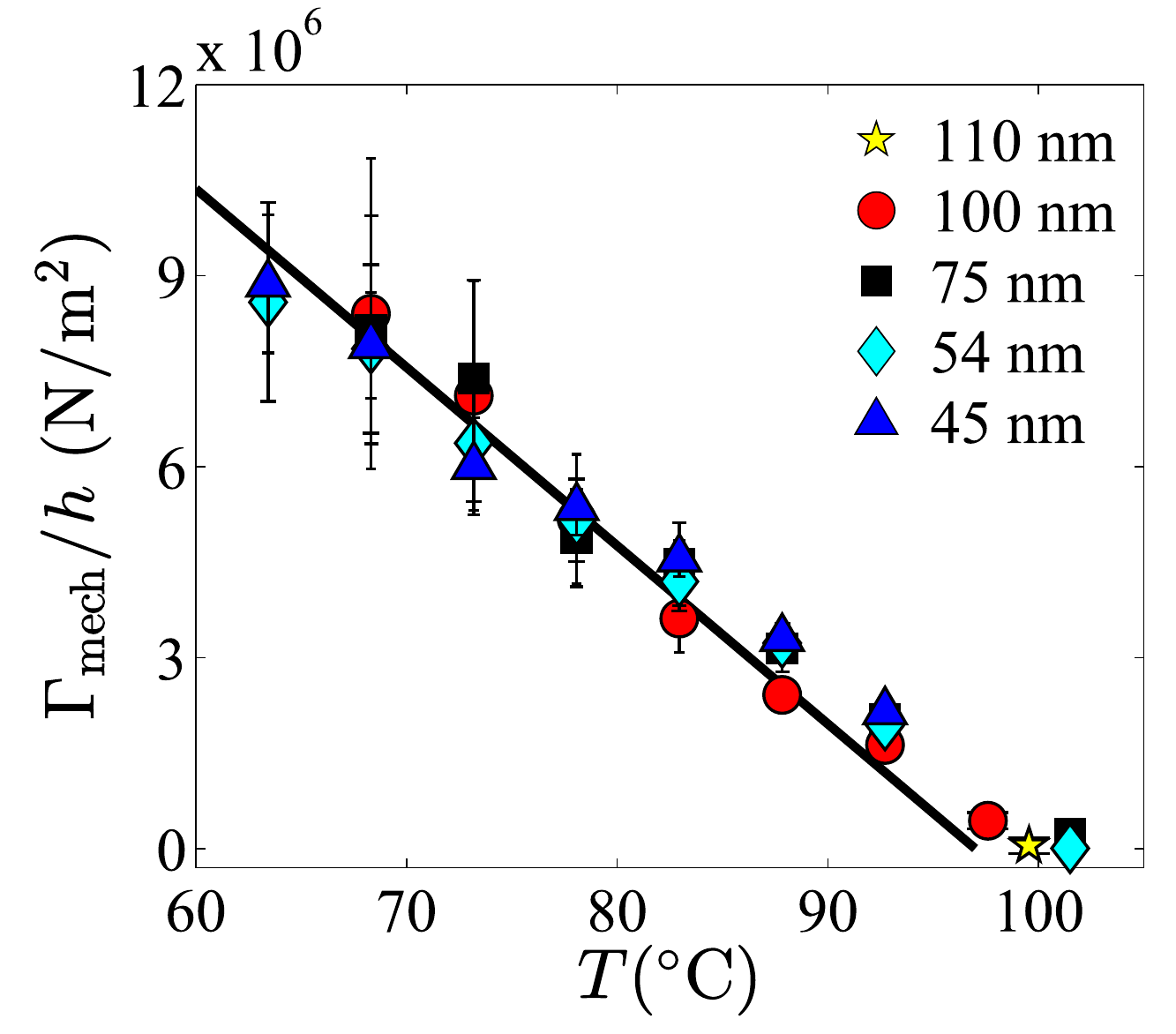}
\caption{Mechanical tension within the film normalized by $h$. A straight line with an $x$-intercept of 97$^\circ$C is fit to the data. }
\label{tension}
\end{figure}

For an elastic membrane subjected to a radial strain, the mechanical tension is equal to $\Gamma_\mathrm{mech} = \sigma h = \epsilon h E / (1-\nu)$, where $\sigma$ is the uniform stress within the film and $\nu$ is the Poisson ratio ($\nu$ = 0.33 for PS~ \cite{Brandrup1999}). Normalizing $\Gamma_{\mathrm{mech}}$ by film thickness should thus collapse our data to a single line. By using equation~(\ref{eq:mech2}), the data from fig.~\ref{sind_sinb} can be plotted as $\Gamma_{\mathrm{mech}}/{ h}$ as a function of temperature, and the data for various film thicknesses are shown to collapse to one line as shown in figure~\ref{tension} consistent with the theory. As explained above, the mechanical tension vanishes in the melt state; thus, we fit a straight line to the data below 97$^\circ$C constrained to pass through the point $\Gamma_\mathrm{mech}/h = 0$  at $T=T_{g} = 97^\circ$~C, which is the glass transition of PS~\cite{Brandrup1999}. The equation of this fit line is given by  $\Gamma_{\mathrm{mech}}/h = (97^{\circ}\mathrm{C}-T)(2.8 \pm 0.1)\cdot10^{5}$~N$^\circ\mathrm{C}^{-1}$ m$^{-2}$.  Comparing the fit line in fig.~\ref{tension} to $\Gamma_\mathrm{mech} = \epsilon h E / (1-\nu)$ we can solve for $E$ using the values for $c_{PS}$, $c_{SS}$, and $\nu$ given above, along with the single fitting parameter from figure~\ref{tension} to determine the modulus of PS. We determine $E$ for our PS film to be $3.4 \pm 0.4$~GPa, which is in agreement with literature values~\cite{Brandrup1999}.

As an aside, we note that in a recent study, taut free-standing films of glassy poly (n-butyl methacrylate)  films were prepared in a similar way and the tension was thereafter determined roughly 15$^\circ$C below its $T_g$ using the same technique~\cite{Schulman2015}. For films with $h \sim 100$ nm, it was found that $\Gamma_\mathrm{in}/\gamma \sim 3$ using glycerol as the test liquid. Using the arguments developed here, along with values for the Young's modulus, Poisson ratio, and thermal expansion coefficient of poly (n-butyl methacrylate), we expect $\Gamma_\mathrm{mech}/\gamma \sim 3$  - consistent with the former study~\cite{cappella2005,askadskiui2003}.

Above, $T_g$, the mechanical tension vanishes and the only contribution to the tension stems from interfacial tensions, which change slowly with temperature in comparison to $\Gamma_\mathrm{mech}$. These considerations are in complete agreement with the data above $T_g$ in figures~\ref{sind_sinb} and \ref{tension}.

\subsection{Contact Angle and Tension}

\begin{figure}[h]
\includegraphics[width=1.0\columnwidth]{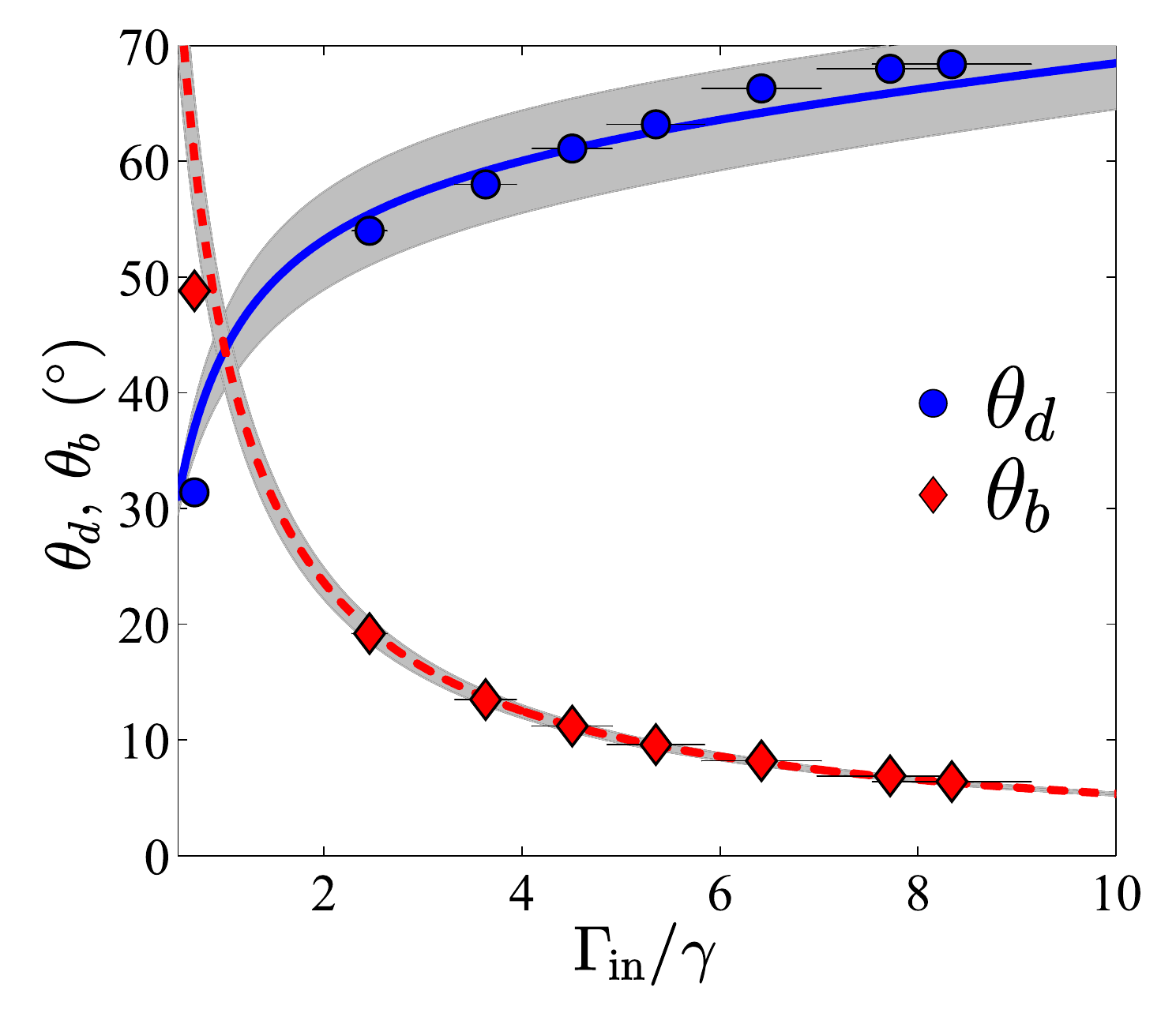}
\caption{$\theta_{d}$ (blue circles) and $\theta_{b}$ (red diamonds) as a function of the film tension in contact with the droplet normalized by $\gamma$ for a film with thickness of 54 nm. The vertical error bars are similar in magnitude to the marker size. Drop and bulge contact angles predicted by the Neumann triangle are shown as solid and dashed lines respectively. Uncertainty in the predicted contact angles are represented by the shaded grey regions and are attributed to the linear fit to $\theta_{y}$ at various temperatures which is used in the calculation of $\theta_{d}$ and $\theta_{b}$. }  
\label{angles}
\end{figure}

Considering the Neumann construction in figure~\ref{fig1}c, we can group the tensions acting in parallel directions at the contact line: in contact with the droplet, inwards, we have $\Gamma_{\mathrm{in}}  =  \Gamma_{\mathrm{mech}} + \gamma_{sv}+\gamma_{sl}$; and away from the droplet in the plane of the film, $\Gamma_{\mathrm{out}} = \Gamma_{\mathrm{mech}} + 2 \gamma_{sv}$~\cite{Schulman2015}. Using the cosine law to perform the tension balance at the contact line, we obtain expression for both $\theta_{d}$ and $\theta_{b}$:
\begin{subequations}
\label{eq:angles}
\begin{eqnarray}
\cos\theta_{d} = \frac{(\Gamma_{\mathrm{out}}/\gamma)^{2} + 1 - (\Gamma_{\mathrm{in}}/\gamma)^{2}}{2 \Gamma_{\mathrm{out}}/\gamma}, \\
 \cos\theta_{b} = \frac{(\Gamma_{\mathrm{out}}/\gamma)^{2} - 1 + (\Gamma_{\mathrm{in}}/\gamma)^{2}}{2 \Gamma_{\mathrm{out}}\Gamma_{\mathrm{in}}/\gamma^2}.
\end{eqnarray}
\end{subequations}
With Young's equation one can show that  $\Gamma_{\mathrm{out}} = \Gamma_{\mathrm{in}}+\gamma \cos\theta_{y}$. Substituting this expression into equations~(\ref{eq:angles}a) and (\ref{eq:angles}b), with the known dependence of $\cos \theta_{y}$ on temperature (fig.~\ref{youngs}) and the known relationship between temperature and $\Gamma_\mathrm{in}/\gamma$  for a given film thickness (fig.~\ref{sind_sinb}), we can predict $\theta_{d}$ and $\theta_{b}$ as a function of $\Gamma_{\mathrm{in}}/\gamma$ and $\theta_{y}$ for the liquid-solid pair. In figure~\ref{angles} we plot the experimentally obtained angles $\theta_d$ and $\theta_b$ as a function of $\Gamma_{\mathrm{in}}/\gamma$  obtained using eq.~(\ref{Eq1}) for a single film thickness. Also shown is the  prediction of eq.~(\ref{eq:angles}a) and (\ref{eq:angles}b), represented by solid and dashed lines. The grey region accounts for the uncertainty in $\cos \theta_y$ determined from figure~\ref{youngs}. 

We now draw attention to the points in figure~\ref{angles} with $\Gamma_{\mathrm{in}}/\gamma = 0.69$. This is a unique set of points because here we observe $\theta_{b} > \theta_{d}$ for the first time. For these data, $T = 101$ $^\circ$C which is above the $T_{g}$ of PS, meaning the measurements were made while the film was in a liquid state. Therefore $\Gamma_{\mathrm{mech}} = 0$ at this point, resulting in $\Gamma_{\mathrm{in}}/\gamma = (\gamma_{sl} + \gamma_{sv})/\gamma$. We can calculate $\Gamma_{\mathrm{in}}/\gamma$ at this temperature using $\gamma_{sl}$, $\gamma_{sv}$, and $\gamma$ from the previous section, and doing so obtain a value of $\Gamma_{\mathrm{in}}/\gamma = 0.7 \pm 0.1$ which is in agreement with our experimental value. 
 
\section{Conclusions}
\label{consec}

In this work, thin glassy films were deformed by the La\-place pressure of sessile liquid drops. The contact angles made by the droplets and corresponding deformations with the undeformed film were measured as tension was varied by tuning film thickness and temperature. Through a Neumann construction at the contact line, it was found that the contact angle of the droplet and bulge could be predicted as a function of temperature and film thickness. We are able to predict the measured tensions, and by comparing to our data to theoretical considerations, the Young's modulus of PS could be obtained. The modulus was found to be in excellent agreement with literature validating the measurements and theory. Furthermore, we show that as the film is raised above the glass transition, the mechanical stress in the film vanishes, and we enter the regime of partial wetting of a liquid on a liquid substrate.

\begin{acknowledgement}The financial support by NSERC of Canada is gratefully acknowledged. The authors thank Ren\'{e} Ledesma-Alonso, Thomas Salez and Elie Rapha\"{e}l for valuable discussions. 
\end{acknowledgement}

%
\bibliographystyle{epj}
\bibliography{pub}
%

\end{document}